# SuSy: a blockchain-agnostic cross-chain asset transfer gateway protocol based on Gravity

A preprint

Aleksei Pupyshev, Elshan Dzhafarov, Ilya Sapranidi, Inal Kardanov, Shamil Khalilov, Sten Laureyssens

Gravity Team

apupyshev@venlab.dev

## Abstract

This document is a specialized technical description of one of the potential implementations of a second layer protocol over Gravity, a blockchain-/token-agnostic decentralized oracle protocol [1].

The SuSy protocol prescribes an implementation of cross-chain transfers of digital assets (tokens) in blockchain networks that support smart contracts, focused primarily on popular blockchains with varying architectures, consensuses and cryptography. SuSy is centered exclusively around technical implementation of transfers, without bringing any incentive models for cross-chain transfer providers.

In addition, we describe the most popular inter-chain communication solutions such as Polkadot, Cosmos Hub, Rainbow and RenVM, as a backdrop for the new solution proposed in this paper.

**Keywords**: oracles, cross-chain, interoperability, interchain transfers, gravity protocol

# Introduction

Gravity is a blockchain-agnostic oracle system that supports communication of blockchains with the outside world, cross-chain communication and sidechains within a single unified structure. The Gravity protocol itself makes no attempt at solving applied problems of inter-chain communication, such as, for example, transferring a token from one blockchain to another. Nevertheless, it is a reliable foundation for such applications, allowing them to remain trustless and decentralized.

Unlike many popular implementations of cross-chain gateways based mainly on Merkle proofs, which is the de facto universal standard for building gateways, we are focusing on a narrower application area in which a gateway operates through a system of smart contracts, verifying only a limited number of clearly specified transactions in both blockchain networks participating in the gateway, with the participation of oracles, which play the role of intermediaries in the transfer of signals from one blockchain to a smart contract of another.

In addition, systems based on Merkle proofs are not capable of resolving the issue of forks within a blockchain. Therefore, in practice, a set of oracles is required, which verify the correctness of a sequence of blocks in blockchain A and report the result to a smart contract of blockchain B. However, such systems are functioning primarily in pairs of blockchain networks that support identical cryptography primitives, working on top of pBFT and/or PoW consensus algorithms.

In this paper, we will briefly review SOTA solutions for cross-chain transfer gateways and describe design requirements of the SuSy protocol running on top of the infrastructure of decentralized oracles represented by the Gravity network.

The presented solution is focused around cross-chain token transfers and makes no attempt to solve general interoperability problems, such as data sharing or cross-chain dApps.

# Solutions

The following section describes popular solutions to the problem of cross-chain gateways in various projects prominent at the time of publication.

# Rainbow

Introduced [2] by NEAR and 1inch project developers, the NEAR <-> Ethereum cross-chain communication protocol is based on a trustless, cryptography-based architecture.

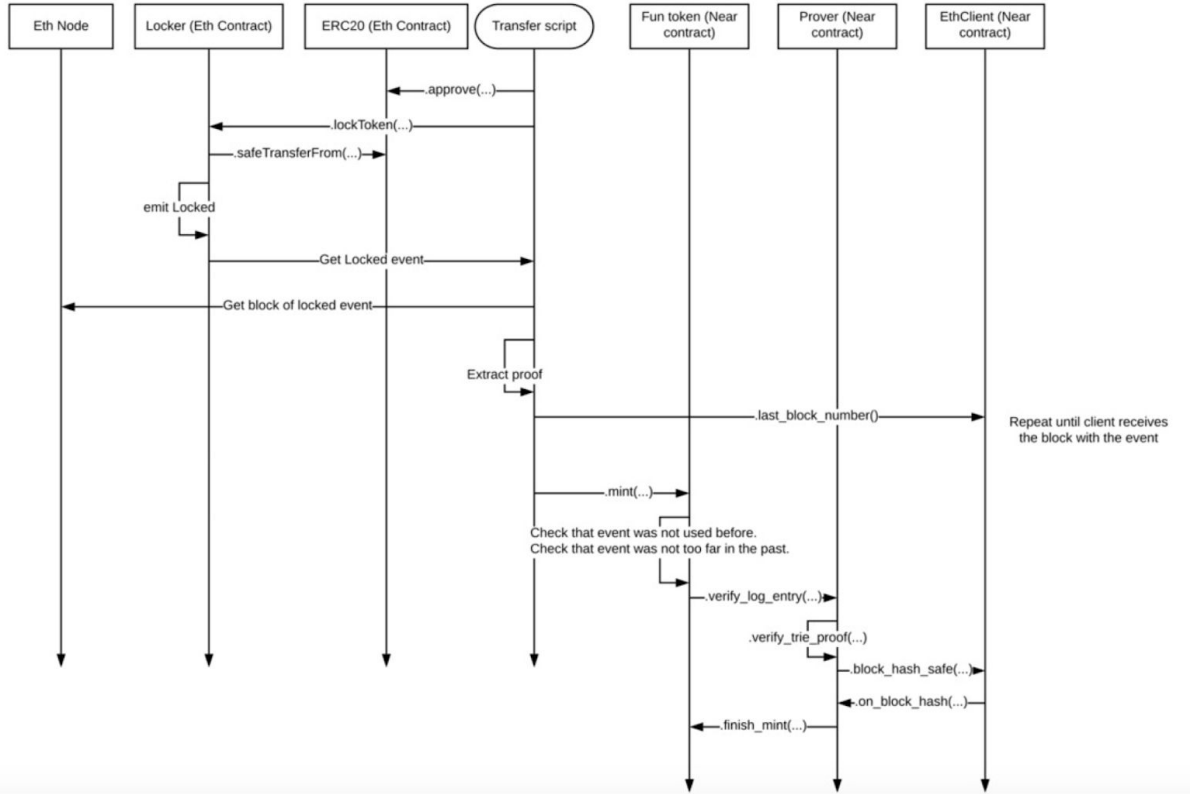

Fig. 1: How smart contracts work in Ethereum and Near networks that implement the rainbow protocol.

The core of the protocol is a feature of smart contracts on NEAR and Ethereum that provides verification of Merkle-proof transactions from the opposite blockchain, and additionally checks the validity of the transmitted chain of blocks. In the Ethereum network, it is verified that there were at least ⅔ validators from the Near network, and in the Near network, the required complexity is checked. The cryptography module verifies a correct sequence of transactions for one chain in another chain's smart contract. The complexity rules and BFT (byzantine fault tolerance) make the system trustless, causing an attack to become as complex and as unlikely as an attack on each of the networks.

When hard forks occur, the gateway smart contracts need to be reinitialized or migrated.

This solution is highly powerful, as it does not require a special set of oracles or validators for gateways, nor a special new utility token to keep it functioning. However, such a solution is not applicable to most blockchain networks in which the consensus mechanism is different from pBFT or PoW, or where cryptography is different.

## RenVM

Ren project [3] represents the most promising class of cross-chain communication algorithms based on Zero-Knowledge (ZK) and secure multi-party computation (MPC) technology, combined with the Byzantine fault tolerant (BFT) consensus mechanism.

Such an architecture, represented by a set of validators (darknodes), allows for cross-chain exchange of any information and for making transfers and other transactions supported by blockchain networks, even those which do not provide smart contract functionality.

All interactions within target chains occur through one single account in each network, and its signature is generated based on shared secrets [4].

The solution is already in operation providing cross-chain transfers between the most popular blockchain networks, Bitcoin and Ethereum. However, it requires a continuing research in the field of cryptography for blockchain networks using different cryptographic primitives. This system also uses the REN utility token, which complicates the model of cross-chain transfers from blockchain A to blockchain B.

## Polkadot

Cross-chain communication in Polkadot is represented by two different mechanics: for the interaction of parachains: specialized internal chains of the polkadot network that are united into a final relay chain, and so-called bridges, for interactions with external popular blockchain networks (for example: Polkadot <-> Ethereum). [5]

The Cross-chain Message Passing (XCMP) protocol is available for communication among parachains and between parachains and a relay chain.

> *"Cross-chain transactions are resolved using a simple queuing mechanism based around a Merkle tree to ensure fidelity. It is the task of the Relay Chain validators to move transactions on the output queue of one parachain into the input queue of the destination parachain. However, only the associated metadata is stored as a hash in the Relay Chain storage.*
>
> *The input and output queue are sometimes referred to in the codebase and associated documentation as "ingress" and "egress" messages respectively. "*
> *- Polkadot Wiki [5]*

In the context of this work, where the primary focus is on the mechanism for cross-chain token transfers between popular blockchains, the usage of bridges [5] is of interest. There are numerous proof-of-concept implementations of such bridges, yet none of those constitute a universal solution. All available bridges offer various approaches to achieving decentralization and communication with external chains, maintaining the unification only in that they are themselves parachains of the Polkadot network.

Despite the fact that Polkadot can be considered as an extremely promising example of a scalable, extensible and flexible protocol, it is still impossible to claim that the project effectively solves vital problems of cross-chain communication between existing blockchain networks. Accordingly, no effective solution for cross-chain token transfers is presented, which would work well in practice.

It is also worth noting that the presence of a separate native token (DOT) in the Polkadot system also significantly limits the system of cross-chain transfers, since it requires purchase / availability of the token to pay network fees.

## Cosmos

The main cross-chain protocol for Cosmos is a protocol developed by the IBC (Inter-blockchain Communication) team. Cross-chain communication in Cosmos is organized through abstractions, such as peg zones and a hub. External blockchain networks interact with peg zones, and communication between them takes place through the hub. [6]

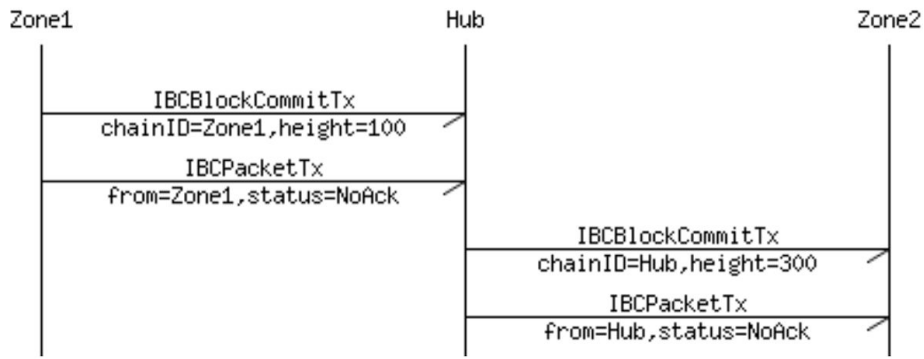

Fig 3.: IBC scheme of cross-chain communication between two blockchains (zones) through a hub. [6]

There are two transaction types:
*IBCBlockCommitTx* is a transaction that allows a blockchain to prove to any observer of its most recent block-hash.
*IBCPacketTx* is a transaction that allows a blockchain to prove to any observer that the given packet was indeed published by the sender's application, via a Merkle-proof to the recent block-hash.

"By splitting the IBC mechanics into two separate transactions, we allow the native fee market-mechanism of the receiving chain to determine which packets get committed (ie acknowledged), while allowing for complete freedom on the sending chain as to how many outbound packets are allowed."

- Cosmos WP [6]

The prerequisite of having the same set of validators in each of the Peg Zones, as necessary in the Cosmos network, is a limitation on the flexibility of this kind of gateway. In particular, this limiting condition raises doubt about the need to use the hub itself as an intermediary in cross-chain transfers.

It is also worth noting that the presence of a separate native token (ATOM) in the Cosmos system significantly limits the system of cross-chain transfers, since it requires purchase/availability of the token to pay network fees.

## SuSy

The SuSy protocol is based on trust in the oracle, which is an intermediary in the transfer of information from one blockchain to another. From a technical standpoint, when implementing the oracle as a trustless decentralized system, which is what the Gravity protocol does, cross-chain gateways on top of it inherit the trustlessness. Another feature of the SuSy protocol implementation over the Gravity oracle protocol is the presence of useful high-level abstractions and services.

Furthermore, for cross-chain swap of a token from one blockchain to another, no additional tokens are required, except for native tokens of the corresponding blockchain networks.

A more detailed description of the SuSY protocol is provided below.

## Key Terms

Let us introduce key terminology to be used in the article:

ORIGIN-CHAIN: a blockchain network from which the transfer originates. That is, in this network, tokens are blocked and unblocked.

DESTINATION-CHAIN: a blockchain, to which transfers are made from the ORIGIN-CHAIN. Issuance and burning of wrapped tokens take place on this network.

sw{TOKEN}: SuSy-wrapped token, a token issued on the DESTINATION-CHAIN blockchain. For example, swETH on networks other than Ethereum mainnet.

IB-PORT is a smart contract in DESTINATION-CHAIN that implements the functionality of issuance and burning of sw{TOKEN}.

LU-PORT is a smart contract in ORIGIN-CHAIN that locks and unlocks the original token.

NEBULA-SC is one of the main architectural units of the Gravity protocol, a smart contract that accepts and verifies data from Gravity oracles. It implements checks of data relevance (blockchain height), availability of appropriate cryptographic signatures and threshold signature rules for transmitted data.

USER-SC is one of the main architectural units of the Gravity protocol. It is a smart contract that accepts data verified in NEBULA-SC and produces an action that is part of a custom application. In the case of SuSy, LU-PORT and IB-PORT are examples of USER-SC.

PULSE-TX is a transaction that will transfer hash from data to NEBULA-SC with necessary signatures for verification and registration.

SEND-DATA-TX is a transaction that transfers data verified and registered in NEBULA-SC to USER-SC.

EXTRACTOR [1] is a Gravity network service that reads and interprets data from external sources. In the case of the SuSy protocol, it reads and interprets data from transactions and states associated with LU-PORT and IB-PORT of the corresponding blockchain networks.

SWAP-ID is a unique identifier for the cross-chain transfer swap operation. For each SWAP-ID, there is a sender in ORIGIN-CHAIN, receiver in DESTINATION-CHAIN, amount and status (registered / processed / finalized). The final status is determined by Gravity oracles, and a rule for setting the status may differ in different blockchain networks.

STATUS-CONTROLLER is a mechanism for managing the status of SuSy-SWAP-ID, the ability to unlock or issue a token if, for some reason, this did not happen in a standard way within a certain time interval (for example, due to high gas fees or blockchain forks).

SUSY-GATEWAY is a set of services and smart contracts that support the transfer of data on the lock/unlock or issue/burn statuses of transferred tokens.

## SuSy Protocol

Let us consider how the T token cross-chain transfer algorithm works using an example of a transfer from ORIGIN-CHAIN to DESTINATION-CHAIN, where it will be issued as a swT token and sent to the recipient R in DESTINATION-CHAIN.

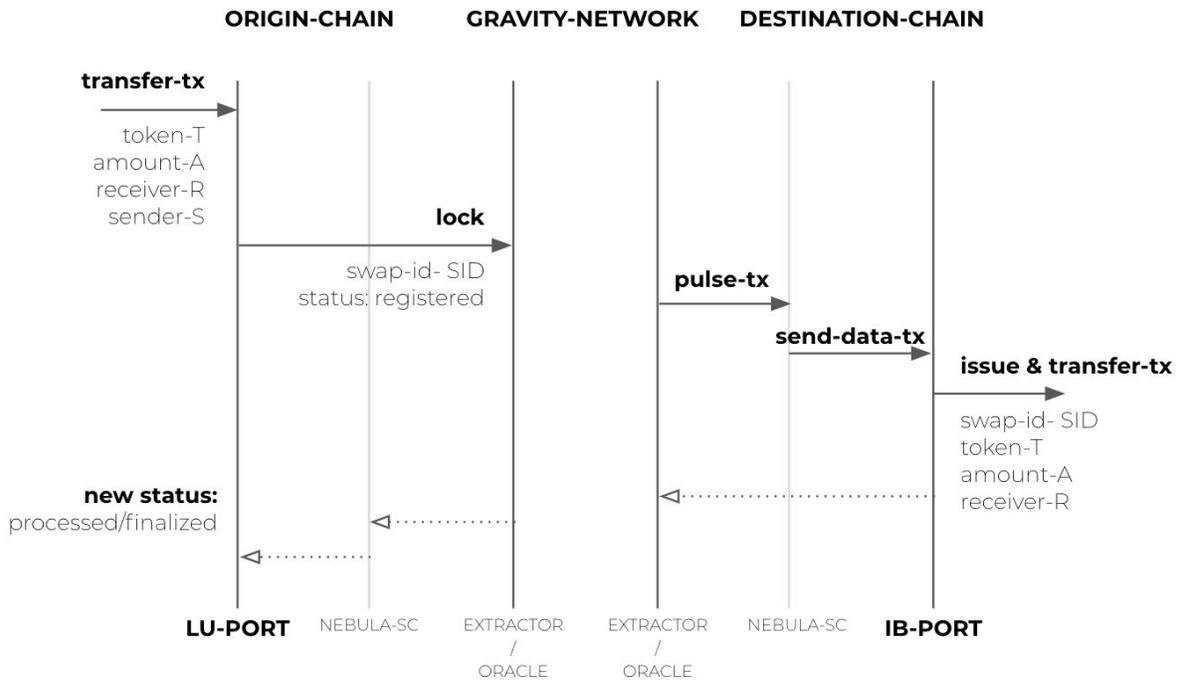

Fig x.: SuSy origin-destination cross-chain transfer scheme.

A user (S) interacts with the LU-PORT smart contract by transferring an amount (A) of the T token to it and specifying the recipient's public address in DESTINATION-CHAIN. The gateway smart contract automatically creates a unique SWAP-ID and sets the registered status. The received funds are blocked on the LU-PORT smart contract.

Information about this event is registered by extractors, the Gravity network's service that processes the received data and communicates it to Gravity. From the Gravity framework, the oracle moves hashed data about the new SWAP-ID and directions to the verification contract (NEBULA-SC), in which the signatures of the Gravity network validators and the legitimacy of the transferred context are checked.

Upon verification, the SEND-DATA-TX transaction is called, containing a set of data and instructions for issuing and sending swT tokens to the recipient (R).

Likewise, data about this event is handled by Gravity network oracles, and, contingent upon successful execution, the "processed" status is set. After reaching a certain number of blocks at which the likelihood of a fork is minimal, it may be necessary to set the finalized status.

In the opposite direction, for transferring the swT token from DESTINATION CHAIN to ORIGIN CHAIN and unlocking T on the LU PORT contract, the procedure is similar.

The only difference is in the final transactions, that is, burning the swT token on IB PORT and unlocking the T token on LU PORT, are reversed.

# Conclusion

The SuSy Gateway protocol scheme described above has a number of advantages and disadvantages. The latter include the necessity for a trustless nature of the oracle network, as well as the non-universality of the protocol that impedes the ability to implement other types of blockchain communications (data exchange triggers etc.) in addition to cross-chain transfers.

Even so, as indicated by the authors, this implementation is quite flexible, straightforward and sufficiently all-inclusive for various blockchain networks supporting smart contracts. The described scheme does not require a special utility token and works via fees paid in native platform tokens.

Additionally, the protocol specification does not describe in any capacity the models of financial incentivization for participants. Yet, it opens up numerous options for creating tokenomics and governance frameworks that can be implemented in accordance with the protocol set forth in this document, keeping it flexible, versatile, scalable to other blockchain networks and expandable to different digital assets.

# References

[1] Pupyshev, Aleksei, et al. "Gravity: a blockchain-agnostic cross-chain communication and data oracles protocol." 2020. arXiv preprint arXiv:2007.00966. https://arxiv.org/abs/2007.00966

[2] NEAR Platform Blog. ETH-NEAR Rainbow bridge. 2020. https://near.org/blog/eth-near-rainbow-bridge/

[3] RenVM. z0 Spec Draft. 2020. https://docs.renproject.io/developers/ z0-spec / z0-spec-draft

[4] Shamir, Adi. "How to share a secret". 1979. Communications of the ACM, 22 (11): 612-613, doi:10.1145 / 359168.359176.

[5] Polkadot Wiki. Bridges. 2020. https://wiki.polkadot.network/docs/en/learn-bridges#docsNav

[6] Kwon, J. and Buchman, E., Cosmos whitepaper. 2019. https://cosmos.network/resources/whitepaper